\documentclass[twocolumn,preprintnumbers,amsmath,amssymb,pre,floatfix,]{revtex4}

\usepackage[]{graphicx}

\usepackage{dcolumn}
\usepackage{bm}

\DeclareTextSymbol{\degre}{T1}{6}
\DeclareTextSymbol{\degre}{OT1}{23}

\begin{document}

\title{Phases of granular segregation in a binary mixture}

\author {Pedro M. Reis$^1$\footnote{email: preis@pmmh.espci.fr}, Tim Sykes$^2$ and Tom Mullin$^2$}
\affiliation{$^1$Physique et M\'ecanique des Milieux H\'et\'erog\`enes, UMR7636 \\ ESPCI/CNRS, 10 rue Vauquelin, 75231 Paris, France  \\
$^2$Manchester Center for Nonlinear Dynamics,\\ University of Manchester, Oxford Road, M13 9PL, UK}

\begin{abstract}

We present results from an extensive experimental investigation into
granular segregation of a shallow binary mixture in which particles
are driven by frictional interactions with the surface of a
vibrating horizontal tray. Three distinct phases of the mixture are established viz; binary gas
(unsegregated), segregation liquid and segregation crystal. Their ranges of existence are
mapped out as a function of the system's primary control parameters
using a number of measures based on Voronoi tessellation. We study
the associated transitions and show that segregation can be
suppressed is the total filling fraction of the granular layer, $C$,
is decreased below a critical value, $C_{c}$, or if the
dimensionless acceleration of the driving, $\gamma$, is increased
above a value $\gamma_{c}$.

\end{abstract}

\maketitle

\section{Introduction}
\label{sec:introduction}

Granular materials, i.e. ensembles of macroscopic discrete
particles, are ubiquitous in our every day life, nature and of
crucial importance in industrial processes
\cite{jaeger:1992,jaeger:1996}. The study of granular media has a
long tradition amongst engineers and geologists who have had
successes in specific problems using a combination of practical
experience and empirical knowledge. More recently the physics
community has taken an increased interest in granular materials
since they pose a number of fundamental questions which challenge
current ideas in non-equilibrium statistical mechanics
\cite{kadanoff:1999}. Interactions between granular particles are
intrinsically dissipative since energy is lost due to both inelastic
collisions and frictional contacts. Hence, any dynamical study of a
granular ensemble requires an energy input which typically takes the
form of vibration or shear \cite{melo:1994,miller:1996}. In this
sense, detailed investigations of granular media provide examples of
canonical systems where dynamical processes  are far from
equilibrium  \cite{egolf:2000}.

An interesting and counter-intuitive feature of particulate matter
is segregation of binary assemblies, where an initially uniform
mixture of particles can spontaneously de-mix into its constituent
components under flow \cite{mullin:2002}. Typically, the 
species of  particles  may differ in size,
density, rigidity or surface properties. Such differences can often
lead to separation and hence clustering of like particles
\cite{williams:1976,bridgewater:1993}. Intriguingly, segregation
does not always happen and the conditions for its occurrence are
difficult to predict. An extensive account of the issues involved
can be found in the following  reviews
\cite{shinbrot:2000,ottino:2000,kudrolli:2004,aranson:2005}.

    \begin{figure}[b]
          \begin{center}
               \includegraphics[width=0.78\columnwidth]{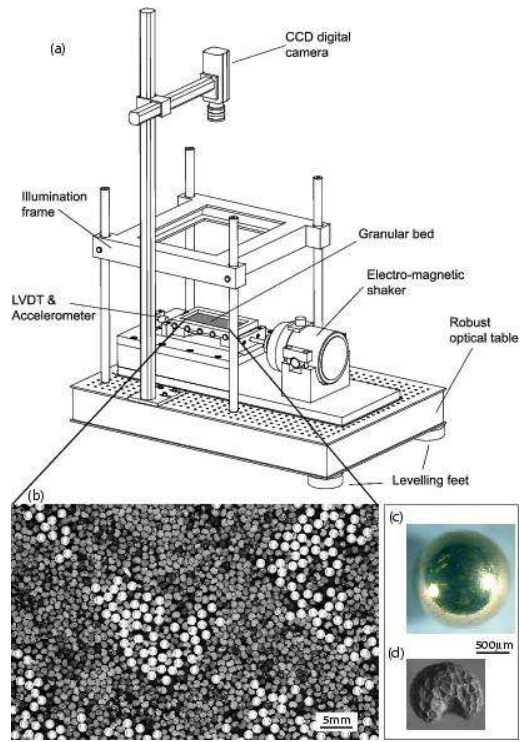}
          \caption{The experimental set up. (a) Three dimensional schematic diagram of the experimental apparatus. (b) Typical segregation patterns of the binary granular mixtures in the segregation liquid phase. Poppy seeds appear as dark grey regions and phosphor-bronze spheres as light gray. The frame was taken after 3min of vibration from an initially homogeneous mixture, ($C=0.708$, $\Gamma=2$, $f=12Hz$, $A=\pm 1.74mm$). (c) Photographs of a phosphor-bronze sphere and (d) a poppy seed. \label{fig:apparatus}}
          \end{center}
     \end{figure}

The phenomena has been recognized for a considerable period of time
\cite{fan:1990} but, despite more than half a century of research,
the underlying nature of the mechanisms involved are not yet
understood. Apart from posing various theoretical questions, insight
into segregation would be beneficial for many applications. These
include areas such as agriculture, geophysics, material science, and
several branches of engineering, e.g. involving preparation of food,
drugs, detergents, cosmetics, and ceramics \cite{fan:1990}. In many
of these examples processing and transporting of mixtures can
lead to undesired separation. Segregation of mixtures can be brought
about by the simple actions of pouring, shaking, vibration, shear
and fluidization and is also found in industrial processes where the
objective is to achieve particle mixing
\cite{bridgewater:1976,williams:1976}.

Over the last decade, significant interest in segregation has arisen in the physics
community. A number of small  scale laboratory experiments have been
reported on vertically \cite{williams:1963, rosato:1987,duran:1993,
shinbrot:1998} and horizontally \cite{betat:1998, painter:2000,
metcalf:2002, aumaitre:2001} vibrated beds, filling and emptying of
vessels \cite{drahun:1983,gray:1997,koeppe:1998,julien:1997} and
rotating cylindrical drums
\cite{donald:1962,clement:1995,gray:1997,hill:1995,choo:1997}. The
spatial distribution and dynamics of segregation of large and small
particles appears to depend on a number of factors besides size
difference including density ratio, friction between particles,
shape of boundaries, particle velocity and the effects of the
interstitial air. A great deal of research is required on the topic  since the parameter space of such a system is
large and seemingly trivial details turn out to have important
effects in the segregation of mixtures.
Designing simple and well controlled laboratory based experiments
and developing appropriate models is therefore essential to make
progress in understand segregation which in turn may give insights
into the industrial problems outlined above.

We have recently developed an experiment to study the
segregation of shallow layers of granular binary mixtures under
horizontal vibration
\cite{mullin:2000,reis:2002,reis:2004,reis:2004b}.  The existence and self-organisation of three phases of
segregation by systematically starting from  homogeneous binary
mixtures have been uncovered. These are, \emph{binary gas} (unsegregated),
\emph{segregation liquid} and \emph{segregation crystal} and they
exist over ranges of the total filling fraction of the layer, $C$.
The principal result is the discovery of critical phenomena in the
segregation process. This implies the existence of a transition
point in $C$ below which the layer remains mixed and above which
segregation occurs.  An overview of this work will be given in
Section \ref{sec:segregation_patterns}.

In this paper we present the results of an extensive experimental
investigation where we show that this phase behavior of the granular
mixture is robust over a range of control parameter space. Moreover,
we investigate the role of the driving on the segregation process
and uncover an additional transition between the segregated and
mixed phases as the dimensionless acceleration of the tray is
increased.  We perform
particle tracking and use the distribution of positions of the
centers of one of the particle types to define a number of measures
derived from Voronoi tessellation to characterize the state of the mixture. This detailed study points to a
robust behavior of our granular mixture which invites the
development of a predictive theoretical model.

This paper is organized as follows. A description of the
experimental apparatus is given in Section \ref{sec:experiment}. In
Section \ref{sec:segregation_patterns} we describe the nature of
segregation patterns observed in our system  and briefly review some
results from our previous work. The microscopic measures obtained
from Voronoi tessellation, namely the local Voronoi density and the
angle between nearest neighbors, are discussed in Sections
\ref{sec:voronoi_density} and \ref{sec:voronoi_angles}. In
Section\ref{sec:parameterspace} we report the results of an exploration of the
parameter space of the system including the aspect ratio of the cell
(Section \ref{sec:parameterspace:aspectratio}), the filling fraction
of the mixture (Section \ref{sec:fillingfraction_parameterspace})
and forcing parameters (Section \ref{sec:parameterspace:forcing}).
Finally, in Section \ref{sec:conclusion} we relate our experimental
results with recent numerical simulations and draw some
conclusions.

\section{The experiment}
\label{sec:experiment}

A schematic diagram of the apparatus is presented in Fig.
\ref{fig:apparatus}a). The experimental set up consisted of a
horizontal rectangular tray with dimensions $(x,y)=(180,90)mm$. It
was connected to an electro-mechanical shaker so that a mixture of
granular particles placed on the tray was vibrated longitudinally.
Individual particles or the granular layer were forced via
stick-and-slip frictional contacts with the oscillating surface of
the container. Different removable frames could be attached to the
tray to make changes in size, shape and aspect ratio.

The tray was mounted on a horizontal platform which was connected to
a Ling LDS V409 electro-mechanical shaker. Its motion was constrained
to be unidirectional by four lateral high precision linear bush
bearings.  The shaker was driven sinusoidally using a  HP 33120A
function generator and the resulting dynamic displacement and
acceleration of the shaking bed were monitored by a  Linear
Displacement Variable Transformer (LVDT) and a PCB quartz shear
piezoelectric accelerometer. 

The main granular mixture consisted of phosphor-bronze precision
spheres and poppy seeds and photographs of representative particles
are shown in Fig. \ref{fig:apparatus}c) and d). The poppy seeds were
non-spherical ("kidney" shaped) with an average diameter of
1.06$mm$, polydispersity of 17\% and a density of $0.2gcm^{-3}$. The
phosphor-bronze spheres had a diameter of $1.50mm$, polydispersity
of 3.0\%  and a density of $8.8gcm^{-3}$. In addition to being
non-spherical the poppy seeds had a considerably larger surface
roughness than the spheres, which resulted in a stronger frictional
interaction with the surface of the oscillating tray. This is
evident from the series of ridges in the surface of the seeds, as
shown in Fig. \ref{fig:apparatus}d)

We define the total filling fraction of the granular layer as,
    \begin{equation}
            C(N_{ps},N_{pb})=\frac{N_{ps}A_{ps}+N_{pb}A_{pb}}{xy}=\varphi_{ps}+\varphi_{pb},
                \label{eqn:compacity}
    \end{equation}
where $N_{ps}$ and $N_{pb}$ are the numbers of poppy seeds ($ps$) and
phosphor bronze spheres ($pb$) in the
layer, $A_{ps}=(0.90\pm0.15)mm^{2}$ and $A_{pb}=1.767mm^{2}$ are the
two dimensional projected areas of the respective individual
particles and $x$ and $y$ are the longitudinal and transverse
dimensions of the rectangular tray. $\varphi_{ps}$ and
$\varphi_{pb}$ are the individual filling fractions for each of the
particle species, respectively. Unless otherwise stated, $C$ was
varied by keeping  the filling fraction of the spheres fixed at
$\varphi_{pb}=0.174$ and changing the filling fraction of the poppy
seeds $\varphi_{ps}$. This enabled more controlled changes to be
made in $C$ but we also investigated changes in both $\varphi_{ps}$
and $\varphi_{pb}$, as will be discussed in Section \ref{sec:fillingfraction_parameterspace}.

In addition to the driving parameters and the total filling
fraction, a geometrical dimensionless parameter, the \emph{aspect
ratio}, is defined as $ \Gamma=\Delta x / \Delta y $ where
$\Delta x$ is the longitudinal length of the tray and $\Delta y$ is
the transverse width of the tray, perpendicularly to the direction
of forcing. The values of $\Delta x$ and $\Delta y$ were changed
using a variable frame which was positioned on all 4 side walls.
Both $\Delta x$ and $\Delta y$ could be independently adjusted to
the required value of $\Gamma$.

All experimental runs were performed in an \emph{approximately
monolayer regime}. The larger heavy phosphor-bronze spheres were
always in a monolayer but the lighter and flatter poppy seeds could
overlap. This degree of overlapping was due to both the
polydispersity of the poppy seeds and the difference in size between
the poppy seeds and the spheres (size ratio of $q\sim0.71$). The
mixture was deemed to be in the a \emph{monolayer regime} if the
extent of overlap of the poppy seeds was never such that the layer
height exceeded the height corresponding to a diameter of the
spheres. The failure of this criterion was readily noticed as
smaller particles were observed to hop over domains of the larger
particles. Hence, for the highest values of the total filling
fraction, $C$ had values higher than those corresponding to maximum
packing in two dimensions, which, for the case of monodisperse disks
is $\pi/\sqrt{12}$. This choice of performing the experiments in
this approximately monolayer regime has two advantages. Firstly,
particles are always in contact with the oscillatory surface of the
tray, such that the forcing was  provided homogeneously throughout
the layer through frictional contacts. Secondly, these approximately
two dimensional experiments allowed the dynamics of the granular layer
to be fully visualized by imaging the system from above.

All experimental runs were started using a homogeneous mixture as
initial conditions. This was achieved using the following procedure.
Firstly, a particular filling fraction of $N_{ps}$ poppy seeds were
vibrated at large amplitude, $A\sim\pm 5mm$ which created a
homogeneous and isotropic layer. The phosphor-bronze spheres were
then suspended above the layer, on a horizontal perforated plate
with ($m\times n$) $2mm$ diameter holes arranged in a triangular
lattice and held by a shutter on an independent superposed frame.
The shutter was then opened and the $N_{pb}=m\times n$
phosphor-bronze spheres fell onto the layer of poppy seeds, creating
a near homogeneous mixture of the two types of particles. An example
of such an initial configuration is presented in Fig.
\ref{fig:experimental_frames}a). We found that if this procedure was
not adopted, random initial clusters in a poorly prepared mixture
could bias the results.

The dynamics of the segregation process were visualized using a near
homogeneous illumination of the tray and the behavior monitored from
above using a CCD camera  as shown in Fig. \ref{fig:apparatus}a).
Phosphor-bronze has a high reflection coefficient compared to the
poppy seeds and the spheres appeared as sharp bright regions so that
direct observation of their motion was relatively straightforward.
The individual positions of the phosphor-bronze spheres were
obtained in a central $(73.1 \times59.5)mm^2$ visualization window
of the full granular layer in order to achieve the necessary
resolution to obtain good estimates of the centers of the spheres.
The spatial distribution of the positions of the phosphor-bronze
spheres were obtained using image processing and particle tracking
techniques which was then used to calculate a variety of measures.
No tracking was carried out on the poppy seeds.

In summary, the control parameters of the experiment were the
amplitude $A$ and frequency $f$ of the applied vibration, the total
filling fraction of the mixture, $C$, and the aspect ratio of the
container, $\Gamma$.

\section{The segregation process}
\label{sec:segregation_patterns}

In our previous work \cite{mullin:2000,reis:2002,reis:2004,reis:2004b} we
present  experimental evidence for three qualitatively distinct
phases of the binary mixture of poppy seeds and phosphor-bronze
spheres. Each exists over a range of  the filling fraction of the
granular layer, $C$. We identify these phases  as \emph{binary gas}
(unsegregated), \emph{segregation liquid} and \emph{segregation
crystal} \cite{reis:2004}.

The \emph{binary gas} phase is found at low values of $C$ and is
essentially a collisional regime. In this phase there is enough free
area and agitation, such that the particles move randomly around the layer and a mixed state persists. In the \emph{segregation liquid} phase, at
intermediate values of $C$, aggregation of the phosphor-bronze
spheres occurs and mobile liquid-like clusters form. A
representative segregation pattern of the mixture in this
segregation liquid phase is shown in Fig. \ref{fig:apparatus}b). The
snapshot corresponds to a segregated state which self-organizes
after the mixture is vibrated for a period of 3 minutes, with
forcing parameters $A=\pm1.74mm$ and $f=12Hz$ and aspect ratio
$\Gamma=2$. The movement of the segregation domains is reminiscent
of oil drops on water. The motion of the particles within the
clusters is highly agitated and the collective motion is slow with
merging and splitting of the domains.

The transition from the \emph{binary gas} to \emph{segregation
liquid} phase \cite{reis:2002} has the characteristics of a
continuous phase transition with square-root dependence of the
saturation levels as measured by the average stripe width which we
treat as an order parameter. This critical phenomenon implies the
existence of a transition point at $C_{c}=0.65$, below which the
layer remains mixed and above which segregation occurs. Moreover,
critical slowing down of the segregation times scales is found near
$C_c$.

At high filling fractions, for $C>0.89$ a second qualitative change
of the structure and dynamics of the domains occurs as well defined
striped patterns, perpendicular to the direction of forcing,
self-organizes from the initial homogeneous mixture.  We denote this
third regime by \emph{segregation crystal}. The monodispersity of
the phosphor-bronze spheres means that the segregation domains in
this phase consist of particles disposed in a hexagonally packed
lattice whereas the polydisperse poppy seeds always move around
randomly.

     \begin{figure}[t]
          \begin{center}
    \includegraphics[width=\columnwidth]{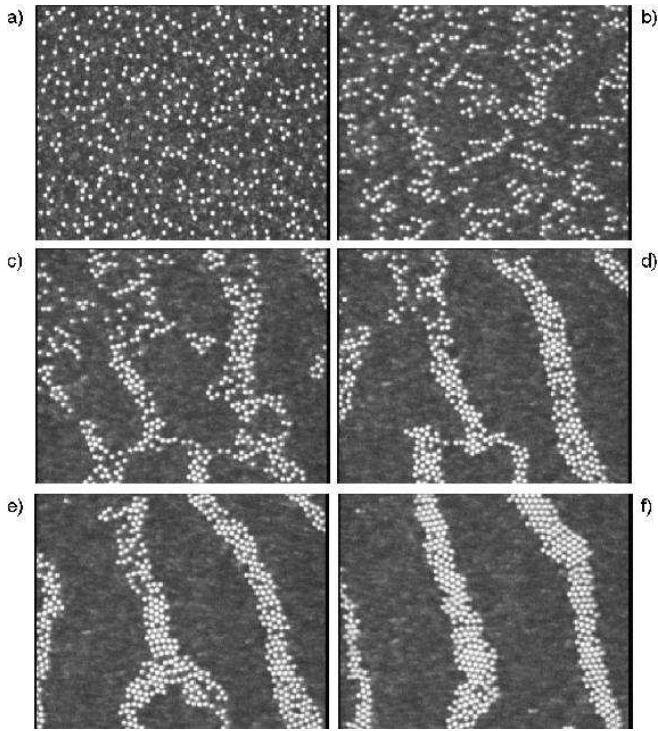}
          \caption{Snapshots of the evolution of segregation domains in a mixture with $C=0.996$, $f=12Hz$, $A=\pm1.74mm$ (a) $t=0s$ -- vibration of the granular layer was initiated from a homogeneous mixture, (b) $t=4.36s$, (c) $t=9.36s$, (d) $t=16.28s$, (e) $t=23.2s$, (f) $t=40.68s$. The snapshots corresponds to a central $(73.1 \times59.5)mm^2$  visualization window of the full tray.
          \label{fig:experimental_frames}}
          \end{center}
     \end{figure}

A  time sequence which is typical of the behavior seen at at high
filling fractions where crystalline stripes form, is presented in
Fig. \ref{fig:experimental_frames}, for $C=0.996$.  Immediately after
the vibration is applied, single large particles diffuse in a sea of
the smaller ones, exploring different local configurations. When two
large particles happen to come close together, the smaller particles
cannot fit between them, and hence the pair is subjected to an
asymmetric pressure that keeps it together. Subsequently, pairs may
encounter others so that progressively larger clusters form. The
unidirectionality of the driving induces an asymmetry in the
segregated domains such that elongated domains of the larger
phosphor-bronze spheres develop in a direction which is orthogonal
to the direction of the drive. During this initial period, the rapid
formation of clusters suggests an effective attractive force between
the phosphor-bronze spheres, that leads to aggregation. Eventually,
long domains form in the y-direction and well defined stripes grow across
the full width of the tray. This self-organization into segregation
domains occurs within timescales of tens of seconds. For longer
timescales of the order of a few hours, these segregated domains
progressively coarsen with time, thereby merging to form
increasingly robust stripes. The width of the domains follows a
$t^{1/4}$ power-law with time and this scaling is independent of the
mixture used \cite{mullin:2000}. The coarsening takes place most obviously for values of $C$ in the range $ \sim$ 0.1--0.2 above $C_{c}$. An extensive parametric investigation of the coarsening behavior as a function of $C$ has yet to be carried out.

It is important to stress once more that, depending on the control
parameters, the mixture does not always evolve into a robust striped
pattern. This was only for filling fractions with
$C>0.89$. At lower filling fractions, in particular closer, but
above, to the segregation transition point $C_{c}$ the segregated
domains are increasingly mobile and blob-like, as shown in the
representative liquid state of Fig. \ref{fig:apparatus}b).


\section{Voronoi Measures}

We now focus on the description of the segregation in our granular
mixture using a number of  \emph{microscopic} measures.  By
microscopic, we mean that both structural and dynamical quantities
are analyzed using the positions of the centers of the individual
phosphor-bronze spheres obtained from the particle tracking
analysis. This is by way of contrast with the macroscopic average
width of the segregated domains and respective fluctuations used in
our previous work \cite{reis:2002,reis:2004,reis:2004b}.

The construction of Voronoi cells through tessellation (also know as
Wigner-Seitz cells) is a standard tool for the study of spatial
configurations  of particle ensembles which is widely used in
condensed matter physics \cite{okabe:1992} and is outlined as
follows. Consider a set $\mathbf{P}$ of $N$ coplanar particles with
their centers located at $C_{i}(x,y)$ for $i=1\rightarrow N$. For
each particle $i$, Voronoi tessellation yields a polygonal cell that
encloses a region inside which any point is closer to the center
$C_{i}$ of the $i^{th}$ particle than any other in the set
$\mathbf{P}$. We have used the \texttt{voronoi(x,y)} routine in the
package MATLAB 7.0 to implement this procedure. It is
straightforward to extract a measure of the local area density
associated with each phosphor-bronze sphere for this geometrical
construction. It is also possible to obtain another useful measure
of the angle between  nearest neighbors. These two quantities are
introduced in the following two Sections, \ref{sec:voronoi_density}
and \ref{sec:voronoi_angles}, respectively.

\subsection{Local Voronoi density}
\label{sec:voronoi_density}

The first quantity considered is a measure of the \emph{local area
density} associated with each phosphor-bronze sphere
\cite{reis:2004}. Following a standard procedure \cite{kumar:2005}, the \emph{local Voronoi area density} of the $i^{th}$ sphere
of an individual video frame can be defined as the ratio,
    \begin{equation}
            \rho^i_v=\frac{A_{sphere}}{A^i_{cell}},
        \label{eqn:voronoi_definition}
    \end{equation}
where $A_{sphere}=\pi(d/2)^2$ is the two-dimensional projected area
of the imaged spheres with diameter $d$ and $A^i_{cell}$ is the area
of its Voronoi polygon.

     \begin{figure}[b]
          \begin{center}   \includegraphics[width=\columnwidth]{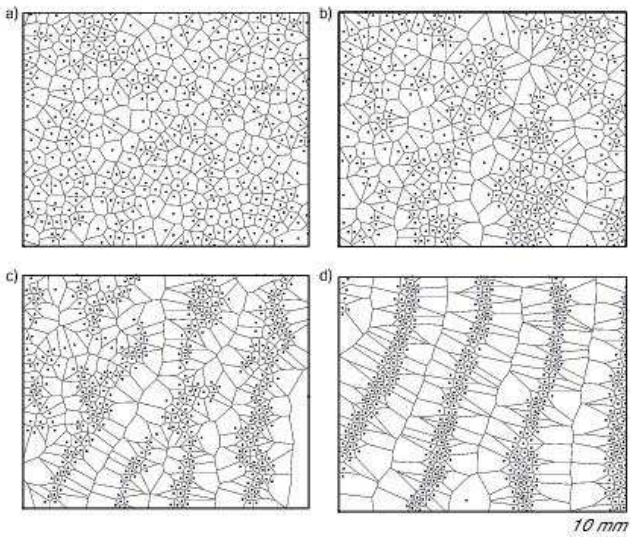}
          \caption{Voronoi diagrams obtained from the positions of the phosphor-bronze spheres, for binary mixtures with various filling fraction values: (a) $C=0.495$, (b) $C=0.623$, (c) $C=0.687$ and (d) $C=0.751$. The frames correspond to configurations obtained $40sec$ after vibrating  an initially homogeneous mixture. \label{fig:voronoi_4examples}}
          \end{center}
     \end{figure}

In Fig. \ref{fig:voronoi_4examples} we present examples of typical
Voronoi configurations, at four different values of $C$, constructed
using the positions of the phosphor-bronze spheres. In  the binary
gas regime at $C=0.495$  no segregation occurs and the network of
Voronoi polygons appear random, as shown in Fig.
\ref{fig:voronoi_4examples}(a). By way of contrast, at $C=1.007$
where definite segregation develops, structure appears in the
domains as can be seen in Fig. \ref{fig:voronoi_4examples}(d). Two
snapshots of the network of Voronoi cells for intermediate values of
$C$ are presented in Fig. \ref{fig:voronoi_4examples}b) and (c).
Note that the particles at the edges of the segregation clusters
have an associated area density significantly lower than those in
the bulk of the domains.

     \begin{figure}[t]
          \begin{center}   \includegraphics[width=\columnwidth]{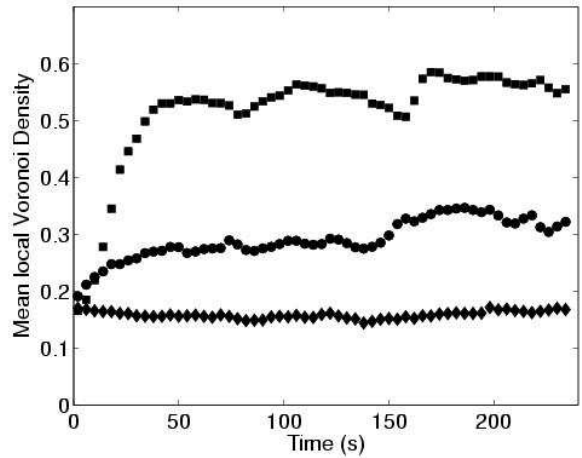}\\
          \caption{(a) Time evolution of mean local Voronoi area density, having started from initially homogeneous mixtures. ({\footnotesize $\blacklozenge$}) $C=0.495$, ({$\bullet$}) $C=0.687$, ({\footnotesize $\blacksquare$}) $C=1.049$.  \label{fig:timedependent_density}}
          \end{center}
     \end{figure}

We first discuss the time evolution of the local Voronoi area
density from the initial mixed state.  A time window of $\Delta
\tau=4sec$, which corresponds to 100 video frames (i.e. $\sim48$
drive cycles) has been used to obtain  dynamic averages for the area
density of individual spheres, $\rho^i_v$, as,
    \begin{equation}
        \overline{\rho_v}\left(t_n=\frac{n\Delta \tau}{2}\right)
        =\langle \rho_v^i\rangle_n,
    \end{equation}
where the brackets $\langle.\rangle_n$ denote averaging over all the
particles, $i$, found within the $n^{th}$ time window, $n\Delta
\tau<t_n<(n+1)\Delta \tau$ with $n \in [0,1,2,3,..,59]$. A time
dependent Probability Distribution Function for the local Voronoi
area density, $PDF(\overline{\rho_v},t_n)$, was obtained by
constructing normalized histograms of $\overline{\rho_v}$ as a
function of the discretised time, $t_n$. Each
$PDF(\overline{\rho_v},t_n)$, for any time window $n$, typically
contained statistical ensembles with 35,000 to
40,000 particles.

In Fig. \ref{fig:timedependent_density} we plot the mean value of
the distribution $PDF(\overline{\rho_v},t_n)$, $D_v(t_n)$, for three
values of the filling fraction. Let us first focus on the behavior at early times, i.e. within
the period $t\lesssim100s$. At $C=0.495$, $D_v(t_n)$ is flat since the layer remains mixed. At
intermediate values of $C$, where $C=0.687$ is a typical example,
$D_v(t_n)$ exhibits a slow increase up to a value of $D_v(t_n)\sim
0.3$ as segregation clusters form. At high values of filling
fraction, of which $C=1.049$ is representative, there is a rapid
initial evolution, since increasingly dense clusters form, up to a
value of $D_v(t_n)\sim 0.525$ after which the mean area density
levels off. This  behavior is consistent with the observation of
saturation in the macroscopic mean stripe width presented in our
previous work \cite{reis:2002,reis:2004,reis:2004b}.

At later times, i.e. $t\gtrsim100s$, the mean Voronoi area density for the two cases of $C=0.687$ and $C=1.049$ only shows small deviations from the level off value of $D_v(t_n)$ and the system has reached a segregated state with a characteristic Voronoi density. The small dips in  $D_v(t_n)$ correspond to the long term splitting and merging of stripes mentioned in Section \ref{sec:segregation_patterns}.  An error estimate of the closeness to a steady state can be obtained from the ratio of the standard deviation to the mean of $D_v(177<t<233 sec)$, which was below $6.5\%$ for all values of filling fraction considered.

   \begin{figure}[b]
          \begin{center}   \includegraphics[width=0.8\columnwidth]{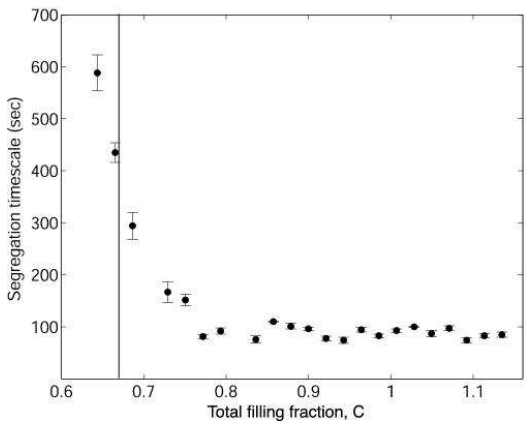}
          \caption{Segregation timescale, $t_D=1/b(C)$ of the phosphor-bronze spheres as a function of filling fraction. The solid vertical line is positioned at the critical point for segregation, $C_c$ obtained in \cite{reis:2004}. \label{fig:densification}}
          \end{center}
     \end{figure}

The dynamical  behavior of the segregation process can now be
analysed as a function of the filling fraction using the local
Voronoi density. We focus on the initial stages of the formation of
domains. During this regime of fast initial segregation growth,
$D_v$ exhibits an approximately linear  behavior of the form,
    \begin{equation}
        D_v(C,t)=D_v(t=0)+b(C).t
        \label{eqn:microscopic:densification_fits}
    \end{equation}
where $D_v(t=0)$ is the mean area density of the initial homogeneous
mixture and $b(C)$ is the corresponding \emph{rate of segregation}.
A value of $D_v(t=0)=0.158\pm0.004$ provides a good fit to all
experimental runs since the layer is consistently started from an
homogeneous mixture. The inverse of the rate of segregation yields a
measure of the segregation timescale, $t_D(C)=1/b(C)$. The quantity
$t_D(C)$ is plotted in Fig. \ref{fig:densification}. The segregation
timescale rapidly increases as $C_c$ is approached from above,
indicating a slowing down of the dynamics which is consistent with
the critical slowing down found previously from the macroscopic
measures  \cite{reis:2002,reis:2004b}.

In \cite{reis:2004} we showed that the  behavior of the
$PDF(\rho_{v})$ distributions obtained in the steady state regime
(i.e. after the initial segregation growth) is  useful in the
characterization of the segregation in a mixture. At low values of
$C$, the PDFs are peaked at small $\rho_v$. As $C$ is increased a
qualitative change in the shape of the PDFs is seen and at
$C\sim0.65$ they flatten out indicating that there is a greater
probability of finding particles with an area density across the
entire range. As $C$ is increased further, a new peak develops at
high area densities corresponding to particles within the
segregation clusters. This peak at high $\rho_v$ becomes
increasingly sharper for high $C$, with a drop at values of
$\rho_v=0.9$, which is consistent with maximum packing in 2D of
$\pi/\sqrt{12}$ for a perfect hexagonal arrangement of disks.

The value of the median of the $PDF(\rho_{v})$ distributions,
$\rho_{v}^{max}$, measures the characteristic Voronoi density of the
spheres in the mixture and is used in Section
\ref{sec:parameterspace} to aid mapping out the phase diagram of the
granular mixture.


\subsection{Angle between nearest neighbors}
\label{sec:voronoi_angles}

\begin{figure}[t]
          \begin{center}   \includegraphics[width=0.8\columnwidth]{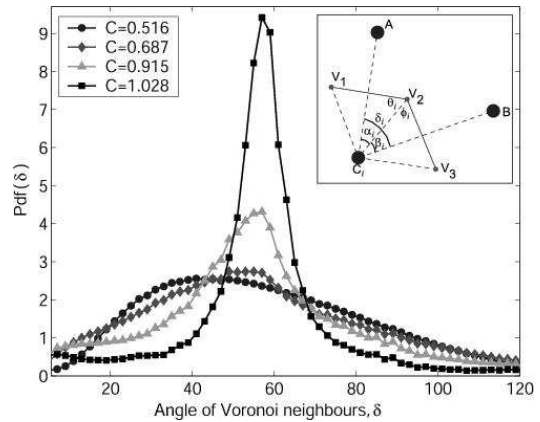}
          \caption{Probability distribution functions for the angle between nearest neighbors, $PDF(\delta)$: for ($\bullet$) $C=0.516$, ($\footnotesize \blacklozenge$) $C=0.687$, ($\footnotesize \blacktriangle$) $C=0.815$, ($\footnotesize \blacksquare$) $C=1.028$. Inset: Schematic diagram for the definition of angles between nearest neighbours, $\delta$. $\mathbf{A}$, $\mathbf{B}$ and $\mathbf{C}_i$ are the positional coordinates of three neighbouring particles. The angle defined by the three particles, about $\mathbf{C}_i$, is $\delta_i=\alpha_i+\beta_i$. The solid lines are a section of the Voronoi polygon, belonging to the  particle $\mathbf{C}_i$, which is defined by the vertices $\mathbf{V}_1$, $\mathbf{V}_2$ and $\mathbf{V}_3$. \label{fig:pdf_delta_4compacities}}
          \end{center}
     \end{figure}

Another quantity that can be calculated from the Voronoi
tessellation procedure  is the angular
distribution between nearest neighbours of each  sphere.
The configuration of a section of a schematic Voronoi polygon for a
particle with coordinates $\mathbf{C}_i$ and two of its nearest
neighbours, $\mathbf{A}$ and $\mathbf{B}$, is given in the inset of
Fig. \ref{fig:pdf_delta_4compacities}. We define $\delta_i$ to be
the angle between nearest neighbours set by $\mathbf{A}$,
$\mathbf{B}$ and $\mathbf{C}_i$. This can be obtained from the
Voronoi tessellation procedure which yields the position of the
coordinates of the vertices of the Voronoi polygon $\mathbf{V}_1$,
$\mathbf{V}_2$ and $\mathbf{V}_3$. Hence, it follows from the
geometric construction around the $i^{th}$ particle that,
        \begin{equation}
            \delta_i=\alpha_i+\beta_i=180-(\phi_i+\theta_i)
        \end{equation}
where,
        \begin{equation}
            \theta_i=\cos^{-1}\left(
            \frac{(\mathbf{V}_1-\mathbf{V}_2).(\mathbf{C}_i-\mathbf{V}_2)}
                 {|\mathbf{V}_1-\mathbf{V}_2||\mathbf{C}_i-\mathbf{V}_2|}
            \right),
        \end{equation}
and         \begin{equation}
            \phi_i=\cos^{-1}\left(
            \frac{(\mathbf{V}_3-\mathbf{V}_2).(\mathbf{C}_i-\mathbf{V}_2)}
                 {|\mathbf{V}_3-\mathbf{V}_2||\mathbf{C}_i-\mathbf{V}_2|}
            \right).
        \end{equation}

Representative curves of $PDF(\delta)$  are
presented in Fig. \ref{fig:pdf_delta_4compacities} for four different values of $C$. For example, for
small $C$ the distribution is broad which indicates that the spheres do not exhibit a preferred orientation as
expected for a mixture. However, as the filling fraction is
increased the main peak of the distribution shifts towards
$60^\circ$ and the width of the peak decreases significantly.

The location, $\delta^{max}$, and width, $w$, of the peak of the
$PDF(\delta)$ distributions is therefore a measure of the local
orientational structure nearby each phosphor-bronze sphere. Both of
these quantities are used in Section \ref{sec:parameterspace} to aid
mapping out the phase diagram of the granular mixture.


\section{Exploration of parameter space}
\label{sec:parameterspace}

     \subsection{Aspect ratio}
     \label{sec:parameterspace:aspectratio}

Since $C$ is a dimensionless ratio between the total projected area occupied by all the particles and the area of the tray, it is now of interest to explore the effect of varying the total area and aspect ratio of the tray. A frame with movable walls in both $x$ and $y$ was used to accomplish this.  Its side walls were mounted in  precision machined parallel grooves and the respective dimensions adjusted using  a Vernier scale which was accurate to within  ±0.05mm. The range of the aspect ratio explored was $0.22<\Gamma<9.00$. The filling fraction of the layer was fixed at $C=0.900$ with $\varphi_{pb}=0.174$ phosphor-bronze spheres. Since the total area of tray varied while changing $\Gamma$, both $N_{ps}$ and $N_{pb}$ had to be changed accordingly in order to keep $C$ and $\varphi$ constant. The forcing parameters, as before, were set at $f=12Hz$ and $A=\pm1.74mm$.

\begin{figure}[t]
          \begin{center}   \includegraphics[width=\columnwidth]{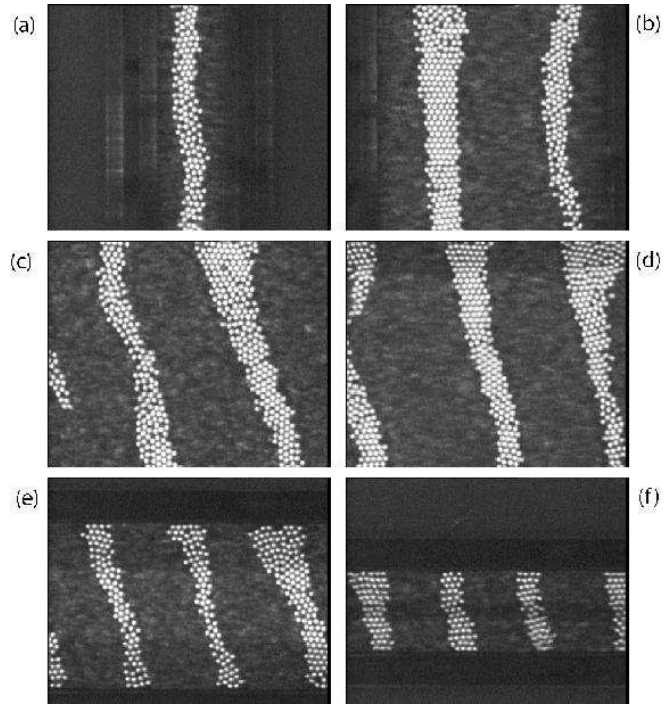}
          \caption{Snapshots of the segregated phase for various values of $\Gamma$. The images correspond to the layer configuration, $2min$ after vibrating an initially homogeneous mixture with the following parameters: $C=0.900$, $\varphi_{pb}=0.174$, $f=12Hz$ and $A=\pm1.74mm$. (a) $\Gamma=0.222$ (b) $\Gamma=0.667$ (c) $\Gamma=1.556$ (d) $\Gamma=2.571$ (e) $\Gamma=4.500$ (f) $\Gamma=9.000$. The end-wall  boundaries can be seen as dark vertical bands  in (a) and (b) and the sidewalls as horizontal bands in (e) and (f). \label{fig:parameterscape:aspectratio_examples}}
          \end{center}
     \end{figure}

Example snapshots of the granular layer taken after $2min$ of vibration of an initially homogeneous mixture are presented in Fig. \ref{fig:parameterscape:aspectratio_examples}(e-f), for various values of  $\Gamma$. Note that the photographs correspond to a central imaging window of the full tray ($73.1\times59.5mm^2$) and hence in  all experimental frames of Fig. \ref{fig:parameterscape:aspectratio_examples} only a portion of the granular layer is shown. The vertical boundaries of the tray are, therefore, only observable in Fig. \ref{fig:parameterscape:aspectratio_examples} (a) and (b)  and the horizontal boundaries  in Fig. \ref{fig:parameterscape:aspectratio_examples} (e) and (f ).

For $\Gamma\ll 1$, of which the frame in Fig. \ref{fig:parameterscape:aspectratio_examples}(a) is a representative example, a single stripe formed  perpendicular to the direction of forcing. For larger values of $\Gamma$, a number of similar well defined vertical stripes rapidly developed in the same way of the particular case of $\Gamma=2$, which was studied in detail in Sections \ref{sec:segregation_patterns}--\ref{sec:voronoi_angles}. Moreover, the number of stripes in the system increased as $\Gamma$ was increased. For example, increasing the tray's aspect ratio from $\Gamma=0.222$ (Fig. \ref{fig:parameterscape:aspectratio_examples}a) to $\Gamma=0.667$ (Fig. \ref{fig:parameterscape:aspectratio_examples}b) resulted on a increase from one to two stripes.

Even if larger number of stripes could be attained by varying $\Gamma$, in all cases the stripes had a similar form i.e. they had approximately the same longitudinal width even if their transverse height (set by $\Delta y$ of the tray) decreased with $\Gamma$. The characteristic local Voronoi area density of the spheres, $\rho_v^{max}$ was measured as a function of $\Gamma$. As the aspect ratio of the tray was changed, the local area density of the phosphor-bronze spheres remained approximately constant (to within $6\%$) with an average value of $\langle \rho_v^{max} \rangle_\Gamma=0.696\pm0.040 $. This suggests that the segregation  behavior appears to be independent of both the total area and aspect ratio of the tray.


     \subsection{Filling fraction parameter space}
    \label{sec:fillingfraction_parameterspace}

\begin{figure}[t]
        \includegraphics[width=0.9\columnwidth]{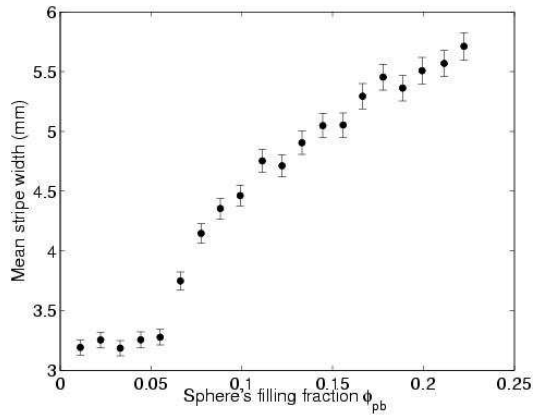}
         \caption{Mean stripe width of spheres as a function of $\varphi_{pb}$. Total filling fraction was kept constant at $C=0.79\pm0.06$. \label{fig:tims_mean_stripe_width}}
\end{figure}

The majority of the experiments were performed by keeping  the filling fraction of the
phosphor-bronze spheres fixed at $\varphi_{pb}=0.174$, and incrementally
increasing the filling fraction of the poppy seeds, $\varphi_{ps}$. Hence it is possible that the transition sequences might be caused by reaching a critical fill ratio of the domain. In order to make sure that this was not the case, a series of experiments were performed where the dependence of the segregation
transition  on the relative composition of the binary
mixture was investigated, while keeping the  $C$ constant. The  amounts of spheres and poppy seeds, set
by $\varphi_{pb}$ and $\varphi_{ps}$, were varied accordingly,
with the constraint of keeping the total filling fraction
of the layer fixed at $C=0.79\pm0.06$. The mean longitudinal
width, $\sigma$, of the domains of phosphor-bronze spheres has been used as the
order parameter to measure the state of the system. Details concerning this measure can be found in
\cite{reis:2002,reis:2004b}. A plot of $\sigma$ versus $\varphi_{pb}$ is given
in Fig. \ref{fig:tims_mean_stripe_width} where it can be seen that for low values
of $\varphi_{pb}$, the mean domain width is
$\sigma=3.23\pm0.04$. This corresponds to approximately two sphere
diameters and is consistent with the chance occurrence of neighboring particles averaged over the layer. Therefore, these are not segregated domains and the layer is in a mixed state. As $\varphi_{pb}$ is increased
past $\varphi_{pb}^{c}$=0.055, $\sigma$ increases continuously. We
emphasize again that $\varphi_{ps}$ has to be changed accordingly in
order to keep $C$ constant. This result suggests that, for
segregation to occur, not only has the total filling fraction  to be
large enough (i.e. above a critical value $C_{c}$ as reported in
\cite{reis:2002}) but also  the amount of spheres in the layer
is required to be above a threshold value $\varphi_{pb}^{c}$. This result emphasizes the fact that the total filling fraction is a two
dimensional parameter $C(\varphi_{pb},\varphi_{ps})$.

Therefore, the dependence of the segregation behavior on $C$ was further explored in  the \emph{filling fraction parameter space},
$(\varphi_{ps},\varphi_{pb})$ by increasing the number of poppy seeds and hence scanning along
$\varphi_{ps}$ for four values of $\varphi_{pb}=\{ 0.054, 0.083, 0.118,
0.174 \}$.
     \begin{figure}[b]       \includegraphics[width=\columnwidth]{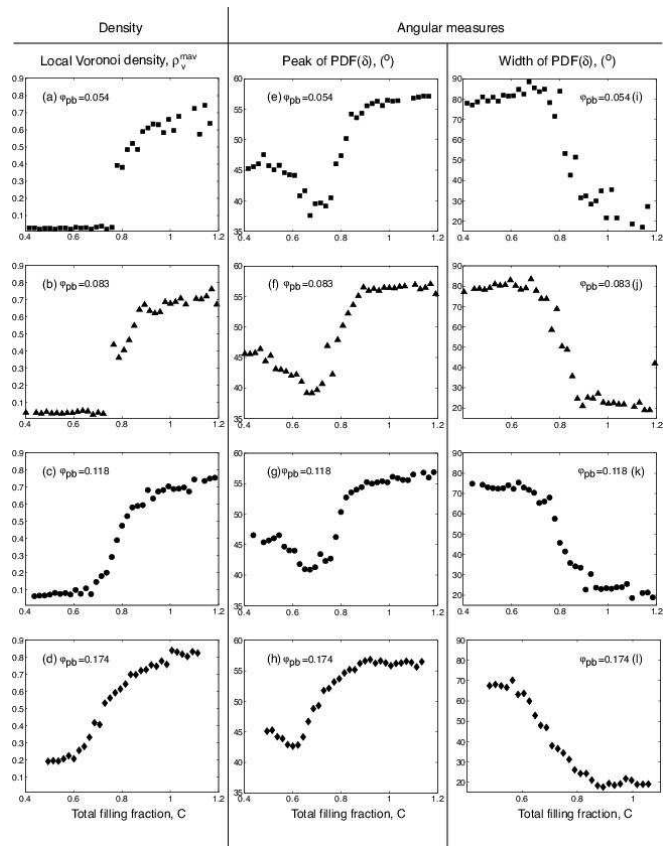}
        \caption{Voronoi measures for the filling fraction phase space. First column: (a--d) Characteristic Voronoi area density. Second Column: (e--h) Characteristic angle between nearest neighbours. Third Column: (i--l) Width of of the peak of the PDF($\delta$) distribution. The x-axis of all plots is the Total filling fraction of the layer, $C$ with $\varphi_{pb}$ set to (a,e,i) $\varphi_{pb}=0.054$, (b,f,j) $\varphi_{pb}=0.083$, (c,g,k) $\varphi_{pb}=0.118$ and (d,h,l) $\varphi_{pb}=0.174$. \label{fig:parameterscape:compacity_voronoi}}
    \end{figure}

The measure used to characterize the state was Voronoi area density,
$\rho_v^{max}$, and the results
are presented in Fig.
\ref{fig:parameterscape:compacity_voronoi}(a-d). All four data sets
exhibit a clear transition between a mixed state (low area density)
and segregated (high area density) state. A particularly interesting
feature is that the transition becomes sharper in the mixtures with
lower values of $\varphi_{pb}$ and yet the
position of the segregation critical point remains fixed. The mean value for the estimate of the mixed state at the left hand edge of each graph is proportionally lower as $\varphi_{pb}$ is decreased. This is expected since the overall area density of the phosphor-bronze spheres in the mixture is  decreased, i.e. the total number of spheres in the tray is smaller for lower $\varphi_{pb}$. However,
the branches corresponding to the segregated phases do not show any significant change, for different $\varphi_{pb}$. Estimates of the  local
densities of $\rho_v\sim0.8$ were attained in all four cases, at
the highest values of $C$.

The location of the maximum of the $PDF(\delta)$ distributions
presented in Section \ref{sec:voronoi_angles} yields the
characteristic angle between nearest neighbors, $\delta^{max}$. In
Fig. \ref{fig:parameterscape:compacity_voronoi}(e-h) we present the
results of $\delta^{max}$ for the four data sets with different
$\varphi_{pb}$, plotted as a function of $C$. Striking  behavior becomes
apparent in this measure for lower values of $\varphi_{pb}$. Take,
for example, the data in Fig.
\ref{fig:parameterscape:compacity_voronoi}e) for
$\varphi_{pb}=0.054$. At the lowest values of $C$, $\delta^{max}$
tends to $45^\circ$. However, as $C$ was increased, $\delta^{max}$
follows a non-monotonic variation, as the mixture goes through the
segregation transition. The characteristic angle first decreases to $38^\circ$ only to then sharply increase past the
segregation transition point. The nature of this non-monotonic dependence of $\delta^{max}$ on $C$ is unclear, even though it appears to be robust and persists in the other three datasets. At the highest values of $C$ the phosphor-bronze spheres self-organize into a nearly hexagonal pack, for which one would expect $\delta^{max}_{hex}=60^\circ$. This is consistent with the value of $\delta^{max}=(57.2\pm2.0)^\circ$ attained in the experiment at high $C$.

     \begin{figure}[t]
          \begin{center}   \includegraphics[width=\columnwidth]{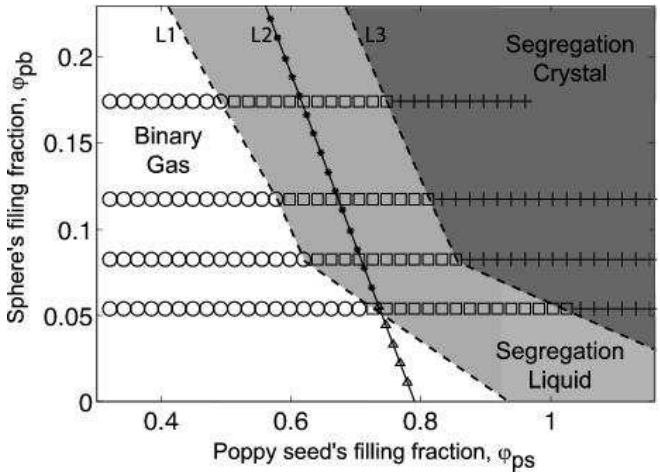}
          \caption{Phase diagram for the $(\varphi_{pb},\varphi_{ps})$ parameter space. ($\circ$ and $\triangle$) binary gas. ($\square$ and $\ast$) segregation liquid. ($+$) segregation crystal. The points along the oblique solid line, $L2$,  correspond to the data presented in Fig. \ref{fig:tims_mean_stripe_width} for which the total filling fraction was kept constant at $C=0.79\pm0.06$ and the relative amounts of particles were changed accordingly (through  $\varphi_{pb}$ and $\varphi_{ps}$). The dashed lines, $L1$ and $L3$ are the experimental phase boundaries between the three phases of the mixture. \label{fig:parameterscape:compacity}}
          \end{center}
     \end{figure}

The second quantity that we extract from the $PDF(\delta)$
distributions is the width at half maximum, $w$, of the peak which is
a measure of the fluctuations in $\delta$ and is plotted
in Fig. \ref{fig:parameterscape:compacity_voronoi}(i-l), for the four values of $\varphi_{pb}$. In all curves,
three distinct regimes are evident where each corresponds to one of the
phases of the binary mixture. As an example, consider the curve for
$\varphi_{pb}=0.118$ given in  Fig.
\ref{fig:parameterscape:compacity_voronoi}(j). At low $C$ the
distribution is broad, indicating that a wide range of angles between
nearest neighbors are possible. This implies that  a disordered
\emph{binary gas} phase exists and there is no segregation. At
$C_c\sim 0.7$, the point at which the segregation transition occurs,
there is a quantitative change of behavior and $w$ rapidly drops
with increasing $C$. This region corresponds to the \emph{segregation liquid} phase. At
$C_m\sim0.9$, $w$ levels off at low values and $PDF(\delta)$ is
now sharply centered at $\delta=60^\circ$, implying that the
transition from the segregation liquid into the segregation crystal
has taken place. This three step behavior in $w$ can be  seen in all four
datasets confirming that the scenario of three segregation phases as
a function of $C$ is robust over a range of the filling fraction
parameter space.

In Fig. \ref{fig:parameterscape:compacity} the four data sets
discussed thus far are combined in a $(\varphi_{ps},\varphi_{pb})$
phase diagram where the phase boundaries $C_c$ and $C_m$ were
calculated from the points demarking the three regimes of $w$
discussed above. The aim is to determine the locus of existence  of the three segregation phases. The binary gas is located
in the left hand side of the diagram in the regions of low
$\varphi_{ps}$. The segregation liquid exists in the central region
of the diagram. The segregation crystal is observed at large values
of $\varphi_{ps}$, on the right hand side of the phase diagram. Note
also that the phase boundaries $L1$ and $L3$ (dashed lines in Fig.
\ref{fig:parameterscape:compacity}) are approximately parallel for
the data sets with $\varphi_{pb}=\{0.083,0.118,0.174\}$, to within
$\sim 9\%$. This indicates that in this region of the phase
diagram the total filling fraction is the primary parameter in
determining the  behavior, rather than the absolute
amounts of each of the particle types. However, larger deviations
occur for the dataset with $\varphi_{pb}=\{0.054\}$. The deviations
in this region of the phase diagram with low $\varphi_{pb}$ is
confirmed by the dataset discussed above where the total filling
fraction was kept constant at $C=0.79$ with $\varphi_{pb}$ and
$\varphi_{ps}$ varied accordingly (data along the oblique solid line
in Fig. \ref{fig:parameterscape:compacity}).

We stress that strong deviations from the results presented here are
to be expected in the limiting cases as the mixture approaches the
single component regimes: i) $\varphi_{pb}\rightarrow 0$ and ii)
$\varphi_{ps}\rightarrow 0$. Two sub-limits of particular interest
are those for the dense cases: iii) ($\varphi_{pb}\rightarrow
\sqrt{12}/\pi,\varphi_{ps}\rightarrow 0$) where crystallization of
the phosphor-bronze spheres prevails and
($\varphi_{pb}\rightarrow 0,\varphi_{ps}\rightarrow 1$) where the
dense liquid-like nature of the poppy seeds  dominates and segregation is 
suppressed. This is consistent with  the polydispersity and shape of the poppy seeds preventing crystallization.


     \subsection{Forcing parameter space}
    \label{sec:parameterspace:forcing}

In the results discussed thus far, the drive parameters have been kept fixed at $f=12Hz$ and $A=\pm1.74mm$.
We now report the results of an investigation of the $(f,A)$ parameter space. The aspect ratio and
total filling fraction were fixed at $\Gamma=2$  and $C=0.900$ with
$\varphi_{pb}=0.174$. All experimental runs in this section, as
before, were initiated from a homogeneously mixed configuration.

     \begin{figure}[b]
          \begin{center}   \includegraphics[width=\columnwidth]{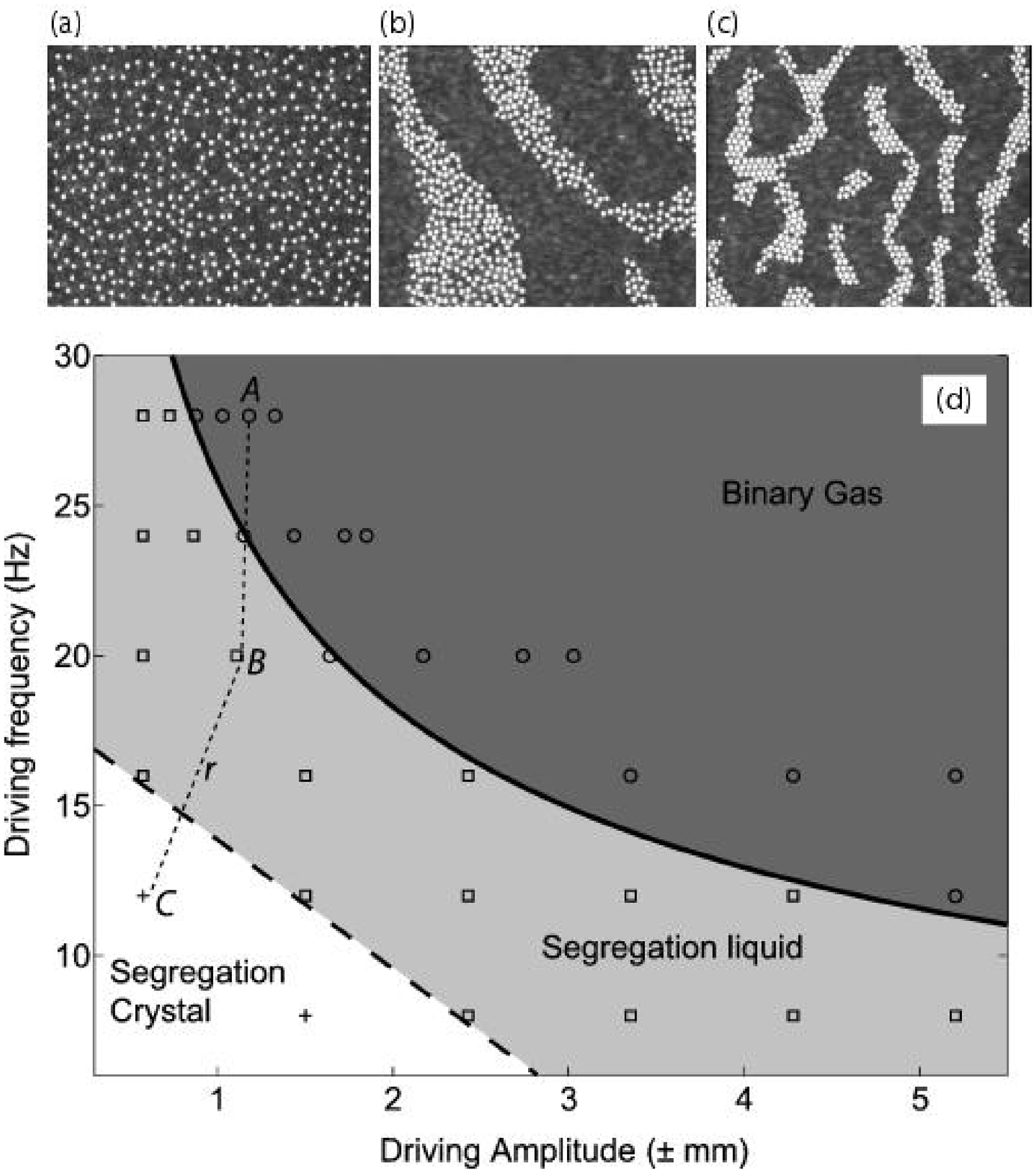}
          \caption{Experimental frames of the binary mixture at various values of $f$ and $A$. (a) Binary gas, $(f,A)$ = $(28Hz,\pm 1.18mm)$. (b) Segregation liquid, $(f,A)$=$(16Hz,\pm 1.50mm)$. (c) Segregation crystal, $(f,A)$=$(12Hz,\pm 0.58mm)$. $\Gamma=2$ ($\Delta y=180mm$) and $C=0.900$ with $\varphi_{pb}=0.174$. (d) Phase diagram for ($f$,$A$) parameter space showing regions of existence for the three phases. ($\circ$) Binary gas. ($\square$) Segregation liquid. ($+$) Segregation crystal. The points $A$, $B$ and $C$, along the parameter path $r$, correspond to the experimental frames shown in (a), (b) and (c), respectively. The phase boundary between the segregation liquid and the binary gas phases is represented by the solid black curve and is given by Eqn. (\ref{eqn:parameterspace:phaseboundary}) with $\gamma_c=2.95.$ \label{fig:freq_amp_phasediagram_withframes}}
          \end{center}
     \end{figure}

With $(f,A)=(12Hz,\pm1.74mm)$ the above conditions produced  a
dense segregation `liquid' from the initial mixture. However, all three phases  can  be obtained at appropriate
locations in the $(f,A)$ space. Examples of each of these
phases can be seen in the experimental frames presented in Fig.
\ref{fig:freq_amp_phasediagram_withframes}(a-c). An example
of a binary gas is shown in Fig.
\ref{fig:freq_amp_phasediagram_withframes}(a) for $(f,A)=(28Hz,\pm
1.18mm)$. The forcing is sufficiently large to induce apparently random motion in the phosphor-bronze spheres  across the layer
and no segregation occurs. This example of a binary gas differs
from the cases considered previously for $C<C_{c}$ in the sense
that, at this high filling fractions ($C=0.900$), the particles in
the layer have persistent contacts with their neighbors and there is
only a small amount of free area in the tray. Therefore this is a highly
agitated but non-collisional state. At $(f,A)=(16Hz,\pm 1.50mm)$, segregation occurred and
the segregation liquid shown in Fig. \ref{fig:freq_amp_phasediagram_withframes}(b) was observed. A representative example of a segregation
crystal is shown in Fig. \ref{fig:freq_amp_phasediagram_withframes}(c)
for  $(f,A)=(12Hz,\pm 0.58mm)$, where spheres within the domains
were disposed in hexagonally packed configurations.

The corresponding phase diagram for the $(f,A)$ parameter space is
presented in Fig. \ref{fig:freq_amp_phasediagram_withframes}d). Two
clear examples of the segregation crystal phase were found at parameter values located in the lower
right hand corner of the diagram i.e. for low amplitudes and
frequencies. At intermediate values of the forcing, a segregation
liquid phase was observed. At relatively large values of $f$ and $A$
segregation did not occur and a binary gas phase was dominant. The
nature of the phase boundary between the segregation liquid and the
binary gas phases (solid line in Fig.
\ref{fig:freq_amp_phasediagram_withframes}d) is discussed below.

In order to obtain quantitative estimates of the behavior, the  $PDF(\rho_v)$ of the distribution of the local Voronoi area density was calculated for each point in  $(f,A)$ space. The
characteristic local Voronoi area density, $\rho_v^{max}$, was
then obtained from the location of the maximum of the distribution, as before.

The dimensionless maximum acceleration of the
tray is used to parameterize the forcing. It is given by ,
        \begin{equation}
            \gamma=4\pi^2\frac{Af^2}{g}
            \label{eqn:parameterspace:gamma}
        \end{equation}
where $g$ is acceleration due to gravity. The parameter $\gamma$ is
commonly used in  vertically vibrated granular systems
\cite{melo:1995} in which gravity plays a dominant role and the
granular layer requires $\gamma>1$ to leave the vibrating base.
In our horizontal set up, gravity enters the problem indirectly
through the frictional forces acted on individual particles,
$F=\mu mg$ where $m$ is the particle mass and $\mu$ its friction
coefficient. In the ideal scenario of no rolling, the value of the
non-dimensional acceleration at which relative motion between the
particle and the oscillatory tray appears would occur at
$\gamma=\mu$.

     \begin{figure}[t]
          \begin{center}   \includegraphics[width=0.8\columnwidth]{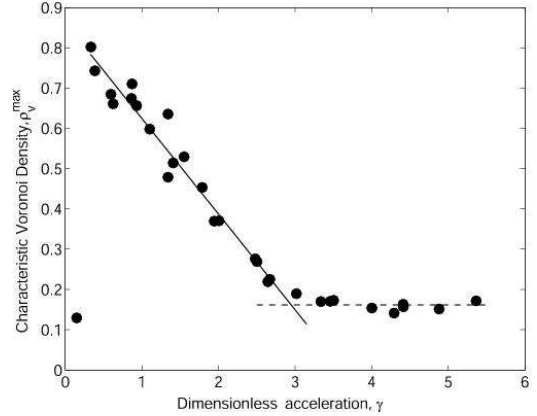}
          \caption{ Plot of the characteristic local Voronoi area density of individual spheres versus the non-dimensional maximum acceleration, $\gamma$ for a range of values of $(f,A)$. The transition is between a mixed (horizontal) and segregated (sloped) states. The solid and dashed lines are the best least squares fit in the segregated and mixed regimes, respectively. The point with the lowest $\gamma$ corresponds to a layer for which the forcing was insufficient to agitate the granular mixture. \label{fig:parameterscape:forcing_transition}}
          \end{center}
     \end{figure}

In Fig. \ref{fig:parameterscape:forcing_transition} the
characteristic local Voronoi area density is plotted as a function
of $\gamma$. All data points from the $(f,A)$ parameter space study
collapse on to  a single curve, where two clearly distinct regimes can
be observed. This indicates that $\gamma$ is, indeed, an appropriate
parameter to describe the forcing of the granular mixture through
oscillatory motion of the tray.

At high accelerations, for $\gamma\gtrsim3$, the
characteristic Voronoi area density remains approximately constant
with increasing $\gamma$. In this regime, the layer was in the
highly agitated binary gas phase discussed above and no
segregation occurred. As $\gamma$ is decreased below $\gamma\sim3$
the final state achieved becomes increasingly more dense in an
approximately linear fashion. The solid and dashed lines in Fig
\ref{fig:parameterscape:forcing_transition} were obtained from
the best least squares fits of the form
$\rho_v^{max}(\gamma)=m\gamma+n$, in the region
$0.34<\gamma<2.67$ (segregation phase) and
$\rho_v^{max}(\gamma)=p$ in the region $3.34<\gamma<5.37$
(binary gas phase), respectively. The intercept of the two
lines yields the location of the transition point which was
measured to be $\gamma_c=2.95\pm0.16$.

The relationship,
    \begin{equation}
        f'=\frac{1}{2\pi}\sqrt{\frac{\gamma_c g}{A'}},
        \label{eqn:parameterspace:phaseboundary}
    \end{equation}
provides a good fit to the phase boundary, $f'(A')$, between the binary gas and segregation liquid phases  as shown by the solid curve in Fig.
\ref{fig:freq_amp_phasediagram_withframes}d). This also
suggests that $\gamma$ is the appropriate parameter to describe
the forcing of the tray.


\section{Conclusion}
\label{sec:conclusion}

We have carried out  a detailed experimental investigation of
segregation behavior in a monolayer of two types of particles contained in a horizontal oscillating tray. The particles are set into motion via stick slip interaction with the base and this effectively randomizes their motion. An initially
homogeneous mixed layer gives rise to spontaneous robust patterns when set into motion. The patterns have the form of
clusters of one particle type and they persist over a wide range of the experimental conditions.  The two principal control parameters of the system
were found to be the layer filling fraction and the dimensionless
acceleration of the driving. Moreover, we demonstrated that the essential mechanisms of the segregation process appeared to
be independent of the aspect ratio of the tray.

Three qualitatively distinct phases were identified: a disordered binary gas for low values of $C$, segregation
liquid with mobile domains of one of the particle types at
intermediate values $C$ and segregation crystal where the domains
are stripes oriented in a direction orthogonal to that of the
driving at high values of $C$. Moreover, we have reported a novel
transition between the segregated and mixed phases as the
dimensionless acceleration of the driving was increased.

Recently, there has been a number of numerical studies where systems
analogous to that of these experiments have been simulated
\cite{pooley:2004,ehrhardt:2005,ciamarra:2005,ciamarra:2005b}. The
segregation behavior of these numerical systems is found to be in
partial qualitative agreement with our experiments.  Ciamarra \emph{et. al.} \cite{ciamarra:2005,ciamarra:2005b} have
suggested that the segregation process is a result of dynamical
instability which resembles the classical Kelvin-Helmholtz
instability observed at a fluid interface. Similar
stripe formation has also been observed in a continuum model of a
binary fluid in which the components are differentially forced
\cite{pooley:2004,sanchez:2004}. These studies suggest  that the
principal mechanism underlying segregation in this class of systems
is the differential driving between each of the particle species. In
our experiments, this could relate to the different surface
properties of the phosphor-bronze spheres and the poppy seeds which
induce different frictional interactions with the surface of the
oscillatory tray.

However, even though the segregation patterns of the simulations are qualitatively similar to those observed in our experiments, none of the numerical models are as yet able to reproduce
the critical behavior. In particular, there is no evidence for the associated
critical slowing down of the dynamics near the transition region. Moreover, our parametric
study of the driving parameters and the additional transition we
have uncovered as a function of the dimensionless acceleration of
the tray demonstrates that the forcing also plays a crucial role in
the phase behavior of the granular mixture. This point has so far
been overlooked in the numerical studies. We believe that this
extensive investigation of the parameter space of our experimental
system will be crucial for further testing and refinement of
numerical models of the type mentioned above.

\emph{Acknowledgments:} PMR was supported by a scholarship from the
Portuguese Foundation of Science and  Technology. The research of TM
is supported by an EPSRC Senior Fellowship. The authors would like
to thank D. Bonamy  for advice on the Voronoi tessellation analysis
and G. Ehrhardt for helpful discussions.

\bibliographystyle{apsrev}

\end{document}